\documentclass{sig-alternate-05-2015}

\begin{document}

\setcopyright{acmcopyright}

\doi{10.475/123_4}



\acmPrice{\$15.00}

%

\title{Ranking Research Institutions Based On Related Academic Conferences}
%
%
%
%
%

\numberofauthors{2} 
%
\author{
%
%
\alignauthor
Yasin Orouskhani\\
       \affaddr{Rahnema Corporation}\\
       \email{y.orouskhani@rahnema.com}
\alignauthor
Leili Tavabi\\
       \affaddr{Intel Corporation}\\
       \email{leili.tavabi@intel.com}
}

\maketitle
\begin{abstract}

The detection of influential nodes in a social network is an active research area with many valuable applications including marketing and advertisement. As a new application in academia, KDD Cup 2016 shed light on the lack of an existing objective ranking for institutions within their respective research areas and proposed a solution for it. In this problem, the academic fields are defined as social networks whose nodes are the active institutions within the field, with the most influential nodes representing the highest contributors. The solution is able to provide a ranking of active institutions within their specific domains. 

The problem statement provided an annual scoring mechanism for institutions based on their publications and encouraged the use of any publicly available dataset such as the Microsoft Academic Graph (MAG)\cite{sinha2015overview}. The contest was focused on research publications in selected conferences and asked for a prediction of the ranking for active institutions within those conferences in 2016. It should be noted that the results of the paper submissions and therefore the ground truths for KDD Cup were unknown at the time of the contest. Each team's final ranking list was evaluated by a metric called NDCG@20 after the results were released. This metric was used to indicate the distance between each team's proposed ranking and the actual one once it was known. 

After computing the scores of institutions for each year starting from 2011, we aggregated the rankings by summing the normalized scores across the years and using the final score set to provide the final ranking. Since the 2016 ground truths were unknown, we utilized the scores from 2011-2014 and used the 2015 publications as a test bed for evaluating our aggregation method. Based on the testing, summing the normalized scores got us closest to the actual 2015 rankings. We therefore used the same heuristic for predicting the 2016 results. 

\end{abstract}

%
%

%
%
\printccsdesc


\section{Introduction}

Influential nodes in social networks are highly sought after in industry and research due to their effectiveness in information diffusion. Finding the influential nodes in a network is therefore an important research area which has many popular applications including viral marketing and advertising. In the research community this topic is valued because of the important role it can play in disseminating new scientific discoveries and technological breakthroughs. 

By the students and academics joining the research community each year, this problem can be viewed as a means of evaluating research institutions within their specific domains. Traditional rankings provided annually by popular newspapers have thus far been used by students to find the most effective institutions. The problem with the traditional rankings is that they come from multiple sources and the used methodology and parameters are often unknown. Therefore the scientific correctness behind these rankings have been in question by some. 

KDD Cup 2016 proposed a solution to this problem by defining each research area as a network consisting of all the active institutions, with the the most influential nodes representing the highest contributors. This method provides an objective ranking measure within specific fields of research according to the institutions' appearances in their respective conferences. The competition was focused on the field of Big Data Mining by looking at selected conferences including but not limited to KDD, SIGIR and ICML in the recent years provided by the Microsoft Academic Graph(MAG). 

The processing of each year's data imposed big data challenges that required the use of Hadoop \cite{White:2009:HDG:1717298} and Spark \cite{Zaharia:2010:SCC:1863103.1863113}. Furthermore each year's conference proceedings amounted to a separate ranking list and therefore a sophisticated rank aggregation technique was needed to provide a final and ultimate  ranking of institutions. Additionally, since the evaluation of the teams was done after the results of the 2016 papers were released, the ground truths were unknown at the time of the contest and this made the competition even more challenging. 

In the following sections, we provide an in depth explanation of our approach. 
In Section 2, we state some of the rank aggregation techniques we tested to provide a background on the methods and in Section 3, we go through the proposed algorithm. The results and conclusions are included in Section 4 and 5 respectively. 

\section{Rank Aggregation Methods}

In order to unify the institutions' rankings computed for each year, we implemented and tested several rank aggregation techniques. In this section, we go over two of these methods to provide an overall background:

\subsection{Borda Count}

Borda count is an intuitive election method in which the voters rank their options in the order of preference\cite{lin2010rank}. The score of each candidate is determined by the total number of candidates standing in the election. Therefore where there are n candidates, a candidate will receive n points for a first preference, n-1 points for a second preference and so on. The scores are then aggregated by a sum or an average to provide a final score list of the candidates which is then sorted to find the ultimate ranking. \cite{de1781memoire}

We describe the algorithm in more detail for the case of full lists. Given full ordered lists $\tau_1$, $\tau_2$, ..., $\tau_L$ each a permutation of the underlying space \textit{T}, we let $ R_{\tau_l}(u) $ be the rank of the element $ u \in T$ in list $ \tau_l $. We let $ B_l(u) $ denote the Borda's score in general, with $ B_l(u) = R_{\tau_l}(u) $ being a special case. Let $ B(u) = f(B_1(u) , B_2(u) , ..., B_L(u)) $ be an aggregate function of the Borda scores. Then one sorts the $B(u)$s to obtain an aggregate ranked list $\tau(T)$. Frequently suggested aggregation functions are:

\begin{equation}
\begin{aligned}
f(x_1, ... , x_L) &= median(|x_1| , |x_2| , ... , |x_L|) (median)\\
f(x_1, ... , x_L) &= (\prod^L_{l=1}|x_l|)^{1/L} (geometric\;mean)\\
f(x_1, ... , x_L) &= \sum^L_{l=1}|x_l|^p/{L} (median)
\end{aligned}
\end{equation}

Note that the method proposed by Borda is a special case of $p$-norm when $p$ = 1 (arithmetic mean) and $ B_l(u) = R_{\tau_l}(u)$, apart from the scaling factor. Although the most frequently used Borda score is the ranking, in situations where additional information is available, the score may be defined accordingly to take other information into account.

\subsection{Fagin Algorithm}

Fagin algorithm is another heuristic method which has two underlying assumptions. One assumption is that individual rank lists are already sorted by their scores and the second assumption is about the monotonicity of the aggregation function.\cite{fagin2002combining}

Given a number of lists, we first perform a sorted access on all lists in parallel and for every item do random access to the other lists to fetch all its values. The process is stopped when we've seen k separate items in the sorted access in all lists. We then sort the list to find the k top items.\cite{ilyas2008survey}

\section{Proposed Algorithm}

In this section, we state our approach for predicting a ranking of the research institutions in 2016. Since the 2016 ground truths were unknown at the time of the competition, we used 2015 publications as a test bed for our methodology. We therefore computed the rank lists from 2011-2014 and used them to propose a ranking for 2015. Comparing our proposed ranking with the actual ranking for 2015, we were able to evaluate our methods and find the one that would result in the closest match.

Our  algorithm uses the institutions' annual scores calculated based on the method provided by the Cup and using the publications retrieved from the MAG. We used all the papers published in the big data mining conferences from 2011-2015 which gave us five separate rankings across the years. 

We then aggregated the rankings to provide a final ranking list for 2016. Here we go through the rank aggregation techniques we tested along with evidence on why our chosen method was the best one to go with. 

Before describing the algorithm, we should introduce some notations. Research institutions are denoted by $u$ and to state the rank of this institution in year $l$, we use $R_{\tau_{l}}(u)$ where $\tau_l$ refers to the ranking list of year $l$. Also to state the score of institution $u$ in year $l$, we use $f_{l}(u)$ and let $f(u) = F(f_{2011}(u) , f_{2012}(u) , ... , f_{L}(u) ) $ be an aggregate function of the algorithms. Consider that $L$ indicates the latest year whose publications have been used, for example in the evaluation step of the algorithm we use $L=2014$ while in prediction phase we set $L$ to 2015.

In the first method, we define a function on the scores of institutions to aggregate rank lists. The score of each institution is obtained from MAG and based on the rules of the Cup. In the next step, we defined a function to sum up the normalized scores of the institutions across the years and sort the resulting list of institutions' scores to create a final ranking. The following function shows the mechanism of this rank aggregation method more explicitly:

\begin{equation}
\begin{split}
f(u) &= F(f_{2011}(u) , f_{2012}(u) , ... , f_{L}(u) ) \\ &= \sum_{l=2011}^{L} f_l(u) / max(\tau_l)
\end{split}
\end{equation}
where $max(\tau_l)$ indicates the maximum value of scores in the rank $\tau_l$.

In the second one, we use the Borda Count method to rank the institutions based on the scores assigned to each position in each list. The scores in this method are the items' positions in the ranking. The score of each position is assigned using the following formula:

\begin{equation}
f_l(u) = R_{\tau_l}(u)
\end{equation}

To create a final rank list, Borda Count adds the scores of each item in all the rank lists and sorts them decreasingly. The following equation describes it mathematically:

\begin{equation}
\begin{split}
f(u) &= F(f_{2011}(u) , f_{2012}(u) , ... , f_{L}(u) ) \\ &= \sum_{l=2011}^{L} f_l(u)
\end{split}
\end{equation}

We also tried the Fagin algorithm as another rank aggregation method which we explained in the second section. 
The following equations provide more intuition:

\begin{equation}
\begin{split}
fu &= F(f_{2011}(u) , f_{2012}(u) , ... , f_{L}(u) ) \\ &= \sum_{l=2011}^{L} f_l(u) / RankListCounts
\end{split}
\end{equation}

where $RankListCounts$ refers to the number of rank lists used.

\begin{table}[H]
	\centering
	\caption{Phase 1 : NDCG@20 Values for SIGIR, SIGMOD, SIGCOMM}
	\label{phase1resulttable}
	\begin{tabular}{|c|c|c|c|} \hline
		{Conf. Name}&Proposed&Borda Count&Fagin\\ \hline
		SIGIR & \textbf{0.823} & 0.74 & 0.80 \\ \hline
		SIGMOD & \textbf{0.876} & 0.724 & 0.712\\ \hline
		SIGCOMM & \textbf{0.713} & 0.703 & 0.649\\ \hline
	\end{tabular}
\end{table}
\begin{table}[H]
	\centering
	\caption{Phase 2 : NDCG@20 Values for KDD, ICML}
	\label{phase2resulttable}
	\begin{tabular}{|c|c|c|c|} \hline
		{Conf. Name}&Proposed&Borda Count&Fagin\\ \hline 
		KDD & \textbf{0.799} & 0.776 & 0.766 \\ \hline
		ICML & \textbf{0.754} & 0.638 & 0.716 \\ \hline
	\end{tabular}
\end{table}
\begin{table}[H]
	\centering
	\caption{Phase 3 : NDCG@20 Values for FSE, MobiCom, MM}
	\label{phase3resulttable}
	\begin{tabular}{|c|c|c|c|} \hline
		{Conf. Name}&Proposed&Borda Count&Fagin\\ \hline 
		FSE & \textbf{0.559} & 0.507 & 0.543 \\ \hline
		MobiCom & \textbf{0.47} & 0.427 & 0.388 \\ \hline
		MM & \textbf{0.394} & 0.349 & 0.349 \\ \hline
	\end{tabular}
\end{table}

\section{Experiments}
\subsection{Experiment Setup}

The algorithm is implemented using Java 1.8 and all experiments were performed on a server running Ubuntu 14.04 with Quad-Core CPU (3.0 GHz) and 8 GB memory. To calculate the scores of institutions, we needed to load the content of a file in the MAG dataset with a size of over 18 GB. Since it was impossible to load the file completely onto the RAM, we developed the proposed algorithm with MapReduce jobs. We therefore chose Apache Hadoop \cite{Vavilapalli:2013:AHY:2523616.2523633} version 2.6  and Apache Spark version 1.6.1 as our implementation platform. 

\subsection{Results}
The goal of KDD cup 2016 was to provide a solution for finding the most influential institutions in several conferences. The competition was run in 3 separate phases, with each phase focusing on different conferences. In this section, we review the results obtained from the experiments in each phase. Phase 1 was performed on SIGIR, SIGMOD, SIGCOMM datasets. Table \ref{phase1resulttable}, shows the NDCG@20 values for each conference in 2015.

As it can be seen in the table \ref{phase1resulttable}, the best method for predicting top institutions in 2015 is our proposed method of adding the normalized scores of each institution from the previous years. This algorithm is employed to solve the task of predicting a ranking of institutions in 2016. After the phase 1 results were announced, this method got an NDCG@20 score of 0.729.

Table \ref{phase2resulttable} illustrates NDCG@20 values for predicting a ranking in the second phase, focused on KDD and ICML. As this table shows, the best method for predicting top institutions in 2015 is the proposed algorithm for KDD and ICML conferences reaching an NDCG@20 score of 0.75.

In the last phase, we needed to predict the top institutions in FSE, MM, and MobiCom conferences. Table \ref{phase3resulttable} provides the NDCD@20 values  for the performance of the algorithm while predicting a ranking for 2015. As shown by the table, our proposed method has a better performance at solving the problem for all the conferences in this phase compared to the techniques mentioned in Section 2. We therefore used the same method for 2016 and got an NDCG@20 equal to 0.74 after the results were announced.

\section{Conclusions}
In this paper, we described our solution to find the most influential research institutions in 2016. The rules of the contest introduced NDCG@20 as a measure of the similarity between each team's ranking and the actual one. We used this metric to find the best ranking mechanism by using the institutions' scores from their previous participation in the conferences.

To predict the highest contributors in a certain academic area by their participation in related conferences, we computed their scores from their previous appearances in the conference and aggregated the scores to reach a final ranking of the institutions. After studying multiple rank aggregation techniques, our conclusion was that adding the scores of the institutions across recent years gets us closest to the true ranking. 

%
\bibliographystyle{abbrv}
\bibliography{sigproc}  
%
%
\end{document}